\documentclass[12pt]{iopart}

\usepackage{graphicx}
\usepackage{bm}
\usepackage{subfigure}
\usepackage[colorlinks=true,linktocpage=true,linkcolor=blue,citecolor=blue]{hyperref}
\usepackage[usenames,dvipsnames]{color}

\usepackage[normalem]{ulem} 
\usepackage{soul} 
\usepackage{nicefrac}

\def\k{{\boldsymbol k}}

\def\dd{{\rm d}}

\def\Peq{{P_{\rm eq}}}
\def\PGZ{{P_\GZ}}
\def\Pmf{{P_{\rm mf}}}

\def\seq{{s_{\rm eq}}}
\def\sGZ{{s_\GZ}}

\def\epseq{{\varepsilon_{\rm eq}}}
\def\epsGZ{\varepsilon_\GZ}
\def\epsmf{{\varepsilon_{\rm mf}}}
\def\epstot{{\varepsilon_{\rm tot}}}

\def\PLtot{{P_\parallel^{\rm tot}}}

\def\PTtot{{P_\perp^{\rm tot}}}

\def\Ieq{{I_{\rm eq}}}

\newcommand{\GZ}{{\rm \scriptscriptstyle{GZ}}}
\newcommand{\onehalf}{{\nicefrac{1}{2}}}

\newcommand{\smallG}{{\rm \scriptscriptstyle{G}}}

\newcommand{\beq}{\begin{eqnarray}}
\newcommand{\eeq}{\end{eqnarray}}
\newcommand{\be}{\begin{eqnarray*}}
\newcommand{\ee}{\end{eqnarray*}}
\newcommand{\bqa}{\begin{eqnarray}}
\newcommand{\eqa}{\end{eqnarray}}

\begin{document}

\title[Kinetics of GZ plasma]{Thermodynamics and kinetics of Gribov-Zwanziger plasma with temperature dependent
Gribov parameter}

\author{Viktor Begun}
\address{Institute of Physics, Jan Kochanowski University, PL-25406~Kielce, Poland}
\ead{viktor.begun@gmail.com}

\author{Wojciech Florkowski}
\address{ $^1$Institute of Physics, Jan Kochanowski University, PL-25406~Kielce, Poland}
\ead{Wojciech.Florkowski@ifj.edu.pl}

\author{Radoslaw Ryblewski}
\address{ $^2$The H. Niewodnicza\'nski Institute of Nuclear Physics, Polish Academy of Sciences, PL-31342 Krak\'ow, Poland}
\ead{Radoslaw.Ryblewski@ifj.edu.pl}

\begin{abstract}
A Gribov-type dispersion relation is treated as an effective
description of interacting gluons forming a hot medium.
Temperature dependence of the Gribov parameter is determined from
the fit to the lattice Yang-Mills results describing thermodynamic
functions. To maintain thermodynamic consistency of the approach,
a temperature dependent bag pressure is introduced.  The
results obtained for equilibrium functions are generalised in the
next step to non-equilibrium conditions. We derive formulas
for the bulk and shear viscosity coefficients within the
relaxation time approximation. We find evidence for largely enhanced bulk viscosity in the region of the phase transition.
\end{abstract}

\pacs{47.10.ad, 24.10.Nz, 47.75.+f}
\vspace{2pc}
\noindent{\it Keywords}: relativistic heavy-ion collisions, quark-gluon plasma, relativistic kinetic theory, Boltzmann equation, relativistic viscous hydrodynamics, shear and bulk viscosities

\maketitle

\section{Introduction}
\label{sect:intro}

Several phenomenological models of the gluon plasma have been constructed in the past~\cite{Goloviznin:1992ws,Peshier:1994zf}, see also \cite{Bluhm:2007cp,Gardim:2009mt,Giacosa:2010vz},  that describe such a system as an ideal gas of massive, noninteracting quasi particles with the dispersion relation of the form
\begin{eqnarray}
E(\k,M) = \sqrt{\k^2 + M^2(T)}. \label{EkM}
\end{eqnarray}
Here ${\bf k}$ is the gluon three-momentum, and $E$ is its energy. A temperature dependent mass, $M(T)$, accounts for the interaction of originally massless gluons.~Thus, in this approximation, the massless interacting gluons become non-interacting massive particles with two possible polarisation directions. We shall continue to call them gluons.

The use of the dispersion relation (\ref{EkM}) in standard thermodynamic expressions for the energy density and pressure
leads, however, to the violation of the basic thermodynamic identities. As it has been pointed out by Gorenstein and
Yang~\cite{Gorenstein:1995vm},  a possible remedy for this situation is the introduction of an additional mean field $B$, commonly
depicted as a bag constant. As the matter of fact, $B$ is not a constant but a function of $M$ and, therefore, it is also a
function of $T$, $B=B(M(T))$. To avoid possible confusion, in the following, we call $B$ the bag pressure.  We note that the name of $B$ originates from the renowned MIT-bag model~\cite{Chodos:1974je}, see also ~\cite{Baacke:1976jv,Shuryak:1980tp,Cleymans:1985wb}.

An alternative description to that using the formula (\ref{EkM}) is the framework based on the Gribov approach to the quantisation of the Yang-Mills theory, that leads to the following dispersion relation~\cite{Gribov:1977wm}
\begin{eqnarray}
E(\k,\gamma_\smallG) = \sqrt{\k^2 + \frac{\gamma_\smallG^4(T)}{\k^2}} \,.
\label{Ek}
\end{eqnarray}
The Gribov parameter $\gamma_\smallG$ is again a function of the
temperature $T$. At high temperatures $\gamma_\smallG$ is known to
depend linearly on $T$. In the region just above the critical
temperature, $\gamma_\smallG$ may be treated as a constant \cite{Zwanziger:2004np,Zwanziger:2006sc}. The
use of Eq.~(\ref{Ek}) has turned out to be quite successful in
qualitative description of thermodynamic properties of a purely gluonic system, 
such as the temperature dependence of the energy density,
$\varepsilon$,  pressure, $P$, entropy density, $s$, and the trace
anomaly (interaction measure),
$I$~\cite{Zwanziger:2004np,Zwanziger:2006sc,Fukushima:2013xsa}.  The Gribov dispersion relation has been also used recently to address real-time problems~\cite{Su:2014rma,Bandyopadhyay:2015wua}.

Clearly, going beyond the approximations that lead to Eq.~(\ref{Ek}) is quite complicated. Instead, in this work we propose to fit the temperature dependence of the Gribov parameter directly to the lattice data~\cite{Borsanyi:2012ve}.  In order to construct a thermodynamically consistent approach, we follow Gorenstein and Yang and introduce a temperature dependent bag pressure.  We should emphasise that the lattice data~\cite{Borsanyi:2012ve} give a strong evidence for the first order phase transition at the critical temperature $T_c~=$~260~MeV, expected for purely gluonic matter. Since our fits use smooth functions, we are able only to determine an approximate behaviour of physical quantities in the phase transition region. The very character of the phase transition, including its order, is not addressed in our considerations.

Having established a thermodynamically consistent model based on Eq.~(\ref{Ek}) we continue to realise our main aim, i.e., we construct a kinetic framework allowing for medium-dependent Gribov parameter. In this way we generalise the results obtained before in Refs.~\cite{Florkowski:2015rua,Florkowski:2015dmm}, where $\gamma_\smallG$ was treated as a constant. In particular, we generalise the formulas for the shear and bulk viscosity.

Determination of the kinetic coefficients for strongly interacting matter is of large interest due to possible applications of such results in hydrodynamic modeling of relativistic heavy-ion collisions~\cite{Torrieri:2007fb,Fries:2008ts,Denicol:2009am,Monnai:2009ad,Bozek:2009dw,Noronha-Hostler:2013gga}. For example, it was argued in~\cite{Torrieri:2007fb} that a sharp rise of the bulk viscosity might cause cavitation and splitting of the plasma into small clusters.
The shear and bulk viscosities were calculated in the high-$T$ regime using, for example, perturbation theory
\cite{Arnold:2000dr,Arnold:2003zc,Arnold:2006fz,Moore:2008ws}, lattice simulations \cite{Meyer:2007dy,Meyer:2007ic}, functional
renormalization techniques \cite{Haas:2013hpa,Christiansen:2014ypa}, quasi particle models~\cite{Sasaki:2008fg,Bluhm:2010qf,Khvorostukhin:2010cw,Plumari:2011mk,Chakraborty:2010fr,Ozvenchuk:2012kh}, and a gas of hadrons~\cite{Chen:2007kx,NoronhaHostler:2008ju}. For conformal systems a lower bound on the shear viscosity to entropy density ratio, $\eta/s \geq 1/4\pi$, was found \cite{Policastro:2001yc}. However, breaking of conformal invariance is necessary for the generation of bulk viscosity \cite{Benincasa:2005iv,Buchel:2005cv}, and a few models have been suggested to reproduce lattice QCD features~\cite{Finazzo:2014cna,Li:2014dsa}.

We note that the bulk viscosity in a gluon system with dispersion relation (\ref{Ek}) was analysed before in a field theoretic approach of Ref.~\cite{Lichtenegger:2008mh} in the context of the old lattice data~\cite{Boyd:1996bx}. Our approach is more phenomenologically oriented and we use the most recent lattice results for gluons~\cite{Borsanyi:2012ve}.  We also obtain qualitatively and quantitatively different results --- the enhancement of the bulk viscosity close to the critical temperature~\cite{Kharzeev:2007wb,Karsch:2007jc}. Such enhancement is particularly interesting due to possible consequences for heavy-ion collisions.

\medskip
The paper is organised as follows: In Sec.~\ref{sect:thermorel} we
discuss thermodynamic relations from the point of view of
thermodynamic consistency and determine the temperature dependence
of the Gribov parameter and  bag pressure.  In Sec.~\ref{sect:kinetics} we
discuss symmetry constraints, the form of the energy-momentum
tensor, the kinetic equation in the relaxation time approximation, 
and the calculation of the bulk and shear viscosities. We
conclude in Sec.~\ref{sect:conclusions}.

\medskip
The three-vectors are denoted by the bold font, the four-vectors are in standard font, the dot denotes the scalar product with the metric \mbox{$g^{\mu\nu}={\rm diag}(+1,-1,-1,-1)$}. The four-vector defining the hydrodynamic flow is denoted by $u$. In the local rest frame (LRF) $u^{\mu}=(1,0,0,0)$.

\section{Thermodynamics}
\label{sect:thermorel}

\subsection{Pressure}

The starting point for constructing a thermodynamically consistent
approach is the definition of equilibrium pressure. The latter is
divided into the particle and mean-field parts, hence we write
\beq
\Peq = \PGZ + \Pmf,
\label{Peq}
\eeq
where
\beq
\PGZ =  - \frac{g T}{(2\pi)^3} \int \dd^3k \ln \left[1 - \exp\left(- \frac{E(\k,\gamma_\smallG)}{T} \right) \right]
\label{PGZ}
\eeq
and
\beq
\Pmf =   - B(\gamma_\smallG).
\label{Pmf}
\eeq
The particle part, $\PGZ$, describes the pressure of the equilibrium
Gribov-Zwanziger (GZ) plasma with the dispersion
relation~(\ref{Ek}). The mean-field part is expressed directly by
the bag pressure. We note, that the Gribov parameter
$\gamma_\smallG$, appearing in the two terms, depends implicitly
on $T$. In Eq.~(\ref{PGZ}) $g=16$ is the degeneracy factor
connected with the internal degrees of freedom (2 (spin) $\times$
8 (color)).  By integrating by parts over momentum in (\ref{PGZ})
we get
\beq
\PGZ &=&    \frac{g}{(2\pi)^3} \int \dd^3k  \, \frac{\k^2}{3 E} \left( 1- \frac{\gamma_\smallG^4}{\k^4} \right) f_\GZ,
\label{PGZ1}
\eeq
where we have introduced the equilibrium (Bose-Einstein) distribution function of GZ gluons
\beq
f_\GZ = \left[\exp\left(\frac{E(\k,\gamma_\smallG)}{T}\right)- 1 \right]^{-1}.
\label{fGZ}
\eeq

\subsection{Entropy density}

The mean field does not contribute to the entropy density, hence we may write
\beq
\seq = \sGZ.
\label{seq}
\eeq
To obtain the particle entropy density we use the Boltzmann definition
\beq
\sGZ = -\frac{g}{(2\pi)^3}  \int \dd^3k \, \Phi[f_\GZ],
\label{sGZ}
\eeq
where the functional $\Phi$ has the form appropriate for bosons, namely
\beq
\Phi[f] = f \ln f - (1+ f) \ln (1+f).
\label{Phi}
\eeq
For $f=f_\GZ$ we find
\beq
\Phi[f_\GZ] = \ln\left[1- \exp\left(-\frac{E}{T}\right)\right]  - \frac{E}{T} \, f_\GZ.
\label{PhiGZ}
\eeq
This gives the equilibrium entropy density in the form
\beq
\seq &=& \frac{g}{(2\pi)^3}  \int \dd^3k \,  \, \frac{E(\k,\gamma_\smallG)}{T}\, {f}_\GZ  \label{seq1} \\
&&  - \frac{g}{(2\pi)^3}  \int \dd^3k \,  \,  \ln\left[1 - \exp\left(-\frac{E(\k,\gamma_\smallG)}{T}\right)\right] \nonumber.
\label{sGZ1}
\eeq
which can be written also as
\begin{eqnarray}
\seq = \frac{2 g}{3 (2\pi)^3} \int \frac{\dd^3k}{E\, T}  \, \left(2 |{\k}|^2 + \frac{\gamma_\smallG^4}{|{\k}|^2}   \right) \,f_\GZ.
\label{seq2}
\end{eqnarray}
The equality of Eqs.~(\ref{seq1}) and (\ref{seq2}) can be verified most easily if one changes to spherical coordinates in the second line of (\ref{seq1}) and integrates by parts over the magnitude of the three-momentum.

In the first line of (\ref{seq1}) we find the energy density of the GZ plasma,
\beq
\epsGZ &=& \frac{g}{(2\pi)^3}  \int \dd^3k \,  \, E(\k,\gamma_\smallG)\, {f}_\GZ,
\label{epseq}
\eeq
whereas in the second line of (\ref{seq1}) we find  $\PGZ$. Therefore, we arrive at the thermodynamic identity
\beq
\seq = \frac{\epsGZ+\PGZ}{T}.
\label{seq3}
\eeq
Since the bag pressure does not appear in (\ref{seq3}),  this formula
can be used to find the temperature dependence of the Gribov
parameter from the temperature dependence of the entropy density
found in the lattice simulations. Similar procedure has been
applied before in the cases where the quasiparticle approach
with the dispersion relation (\ref{EkM}) has been used in
hydrodynamics, for example, see~\cite{Romatschke:2011qp,Alqahtani:2015qja}.

\begin{figure}[t]
\begin{center}
\includegraphics[angle=0,width=0.6\textwidth]{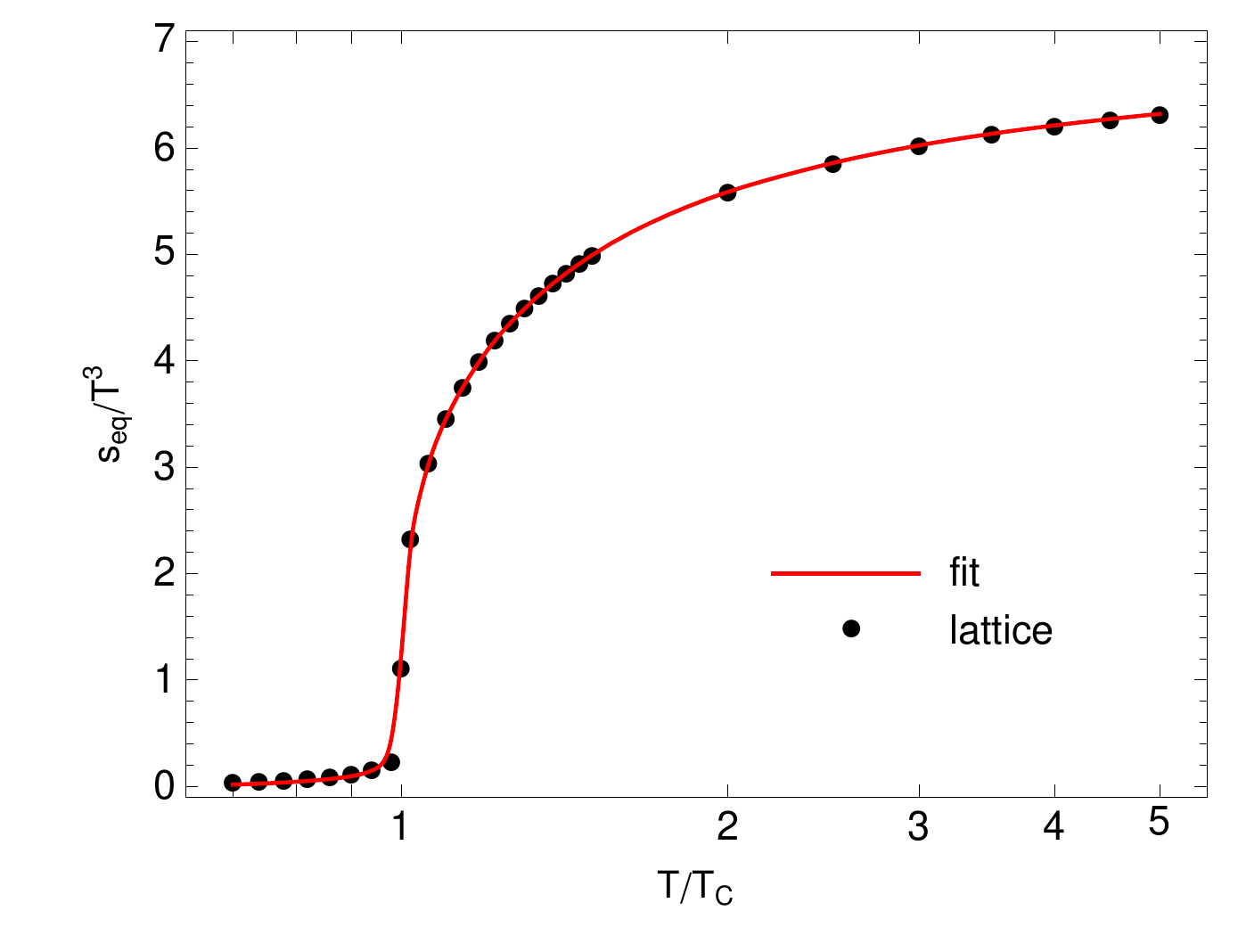}
\end{center}
\caption{(Color online) Temperature dependence of the entropy density scaled by~$T^3$. The lattice data~\cite{Borsanyi:2012ve} (dots) is compared with our approximation (solid red line) based on Eq.~(\ref{sGZ1}) with the function $\gamma_\smallG(T)$ shown in Fig.~\ref{fig:gamma}. }
\label{fig:entropy}
\end{figure}
\begin{figure}[t]
\begin{center}
\includegraphics[angle=0,width=0.6\textwidth]{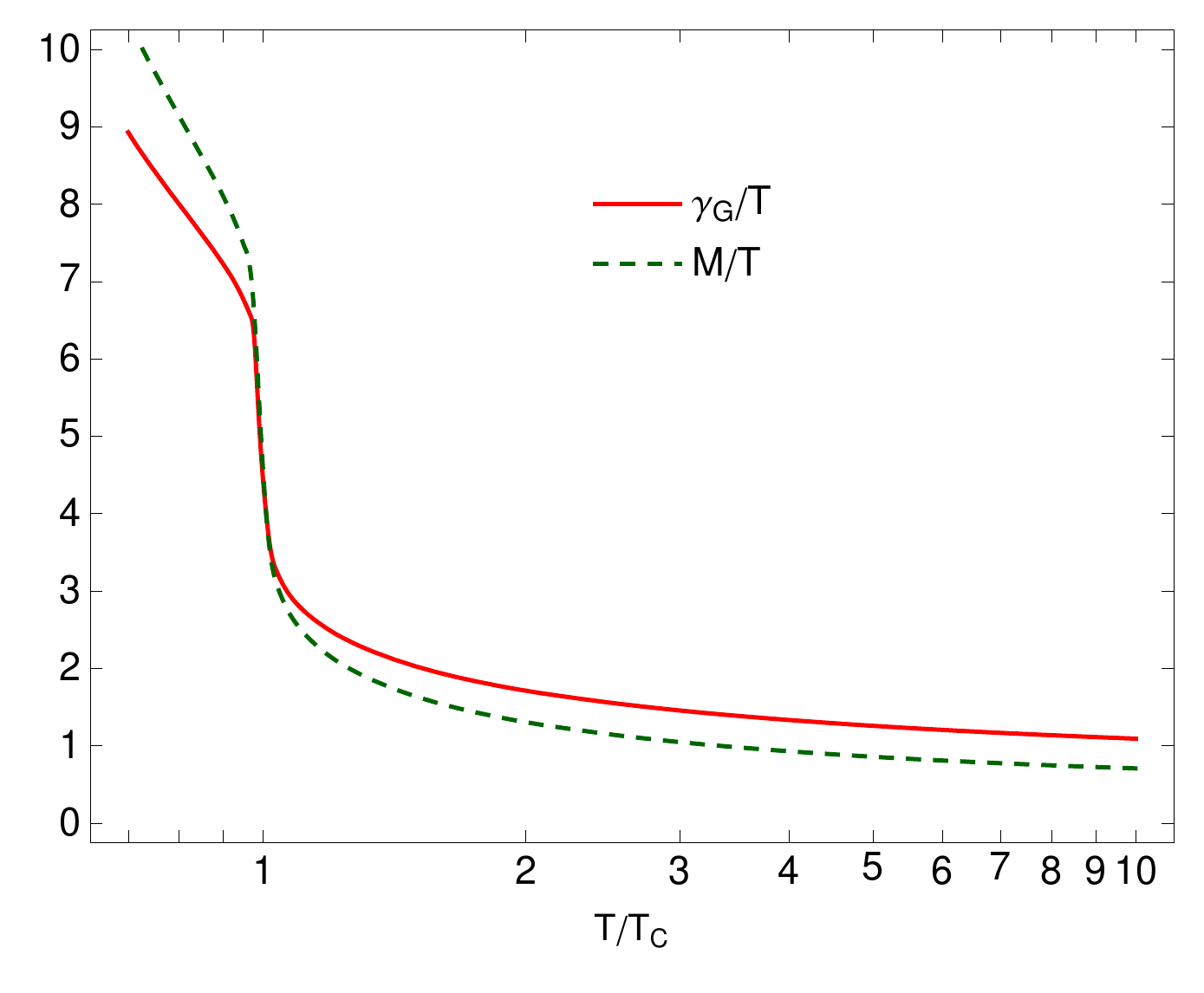}
\end{center}
\caption{(Color online) Temperature dependence of the Gribov parameter $\gamma_\smallG$  obtained from the fit to the lattice results for the entropy density (solid red line). The short-dashed green line represents $M(T)$ obtained in the similar way from the lattice data but with the dispersion relation (\ref{EkM}). }
\label{fig:gamma}
\end{figure}
\begin{figure}[t]
\begin{center}
\includegraphics[angle=0,width=0.6\textwidth]{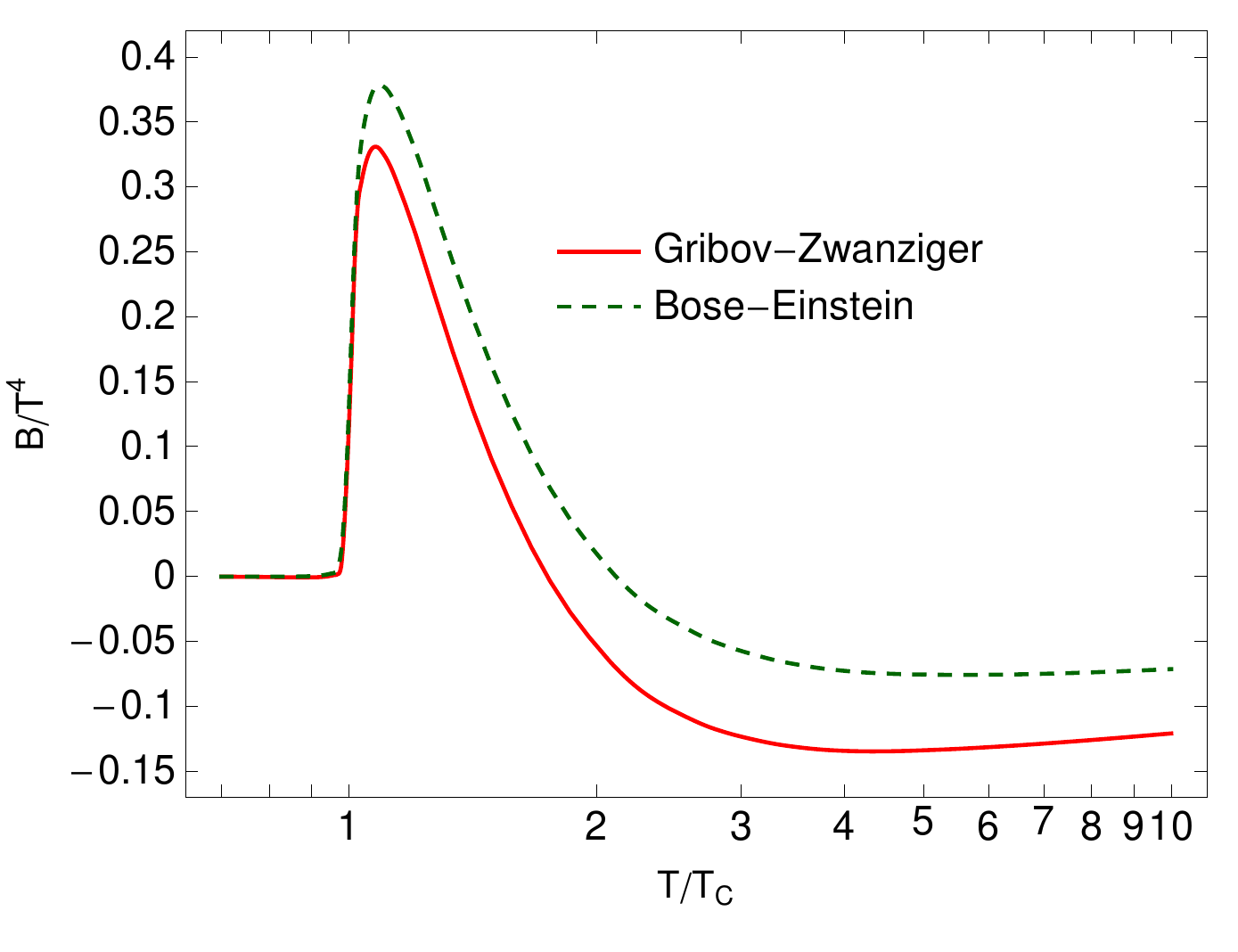}
\end{center}
\caption{(Color online)  Temperature dependence of the bag
pressure obtained from the condition (\ref{Delta1})  that
guarantees thermodynamic consistency of our approach with the
boundary condition $B(T=0)=0$ (solid red line). The short-dashed
green line gives $B(T)$ obtained similarly with the dispersion
relation (\ref{EkM}).} \label{fig:bag}
\end{figure}

\subsection{Energy  density and thermodynamic identities}

Given $\Peq$, the formula for the energy density $\epseq$ is obtained from the thermodynamic identities
\beq
\epseq = T \seq - \Peq, \quad \seq = \frac{\dd\Peq}{\dd T} .
\label{eps1}
\eeq
In our case, using Eq.~(\ref{Peq}) we obtain
\beq
\epseq = \epsGZ + B - T \Delta \frac{\dd\gamma_\smallG}{\dd T}
\label{eps2}
\eeq
and
\beq
\seq = \frac{\epsGZ+\PGZ}{T}  - \Delta \frac{\dd\gamma_\smallG}{\dd T},
\label{seq4}
\eeq
where we have introduced the notation
\beq
\Delta =  \frac{g}{(2\pi)^3}  \int \dd^3k \,  \, \frac{2 \gamma_\smallG^3}{\k^2 E(\k,\gamma_\smallG)}\, {f}_\GZ
+ \frac{\dd B}{\dd\gamma_\smallG}.
\label{Delta}
\eeq
Comparing Eqs.~(\ref{seq3}) and (\ref{seq4}) we conclude that the thermodynamic consistency is maintained only if $\Delta = 0$ or, equivalently, if
\beq
 \frac{g}{(2\pi)^3}  \frac{\dd\gamma_\smallG}{\dd T} \int \dd^3k \,  \, \frac{2 \gamma_\smallG^3}{\k^2 E(\k,\gamma_\smallG)}\, {f}_\GZ
+ \frac{\dd B}{\dd T} = 0.
\label{Delta1}
\eeq
If the temperature dependence of the Gribov parameter is
determined by the fits of the lattice entropy density,
Eq.~(\ref{Delta1}) can be used in the next step to determine the
temperature dependence of the bag pressure. Thus, in the
case $\Delta = 0$ we obtain the compact expression for the energy
density
\beq
\epseq = \epsGZ + \epsmf, \quad \epsmf = B(\gamma_\smallG).
\label{eps3}
\eeq
The expressions for the energy density and pressure can be used to find the interaction measure (also known as the trace anomaly)
\begin{eqnarray}
\Ieq = \epseq - 3 \Peq = \epsGZ - 3 \PGZ + 4 B.
\label{I}
\end{eqnarray}

\begin{figure}[t]
\begin{center}
\includegraphics[angle=0,width=0.6\textwidth]{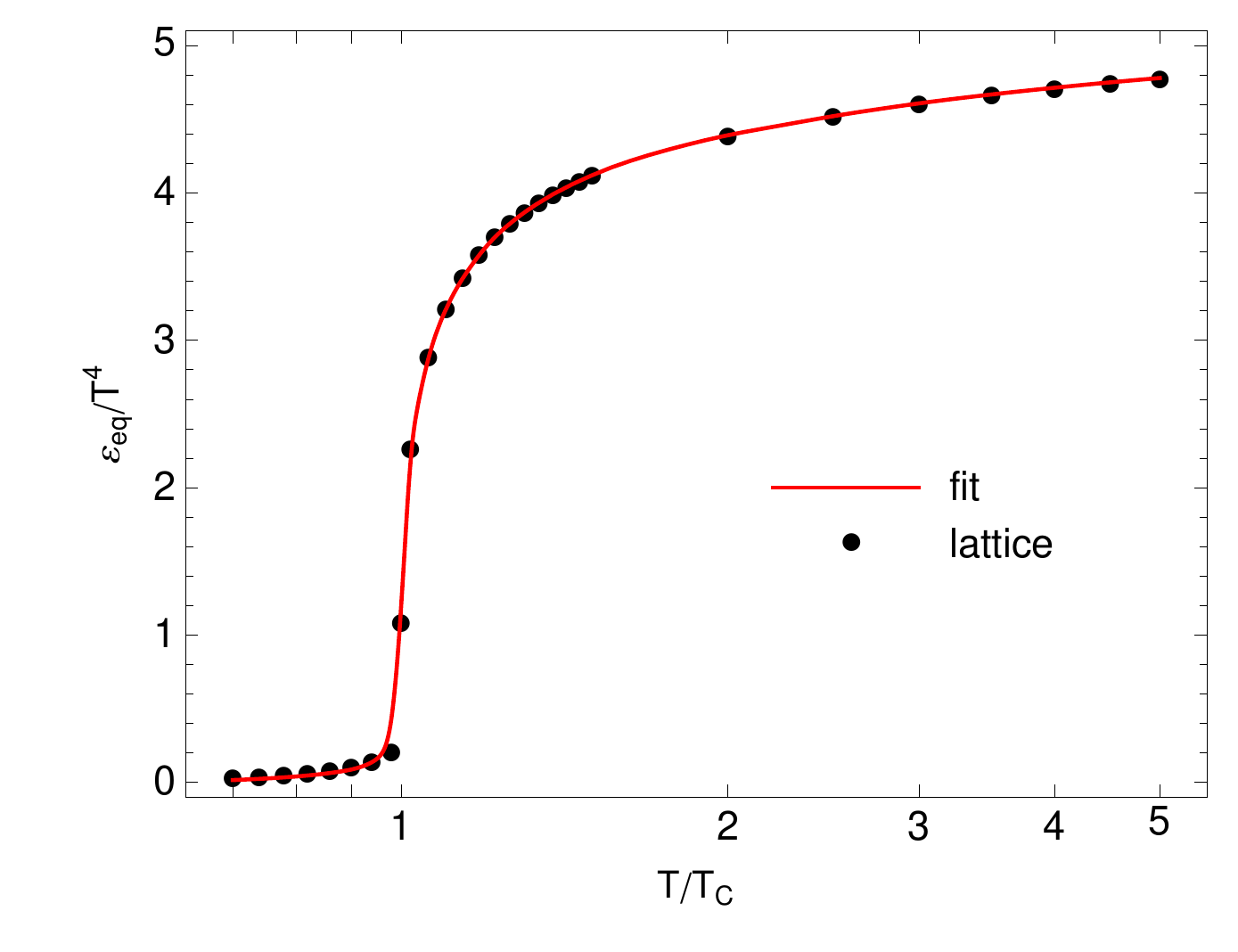}
\end{center}
\caption{(Color online) Temperature dependence of the energy density scaled by $T^4$. The lattice data~\cite{Borsanyi:2012ve} (dots) are compared with our approximation based on Eq.~(\ref{eps3})  with the functions $\gamma_\smallG(T)$ and $B(T)$ depicted in Figs.~\ref{fig:gamma} and \ref{fig:bag}.}
\label{energy}
\label{fig:energy}
\end{figure}
\begin{figure}[t]
\begin{center}
\includegraphics[angle=0,width=0.6\textwidth]{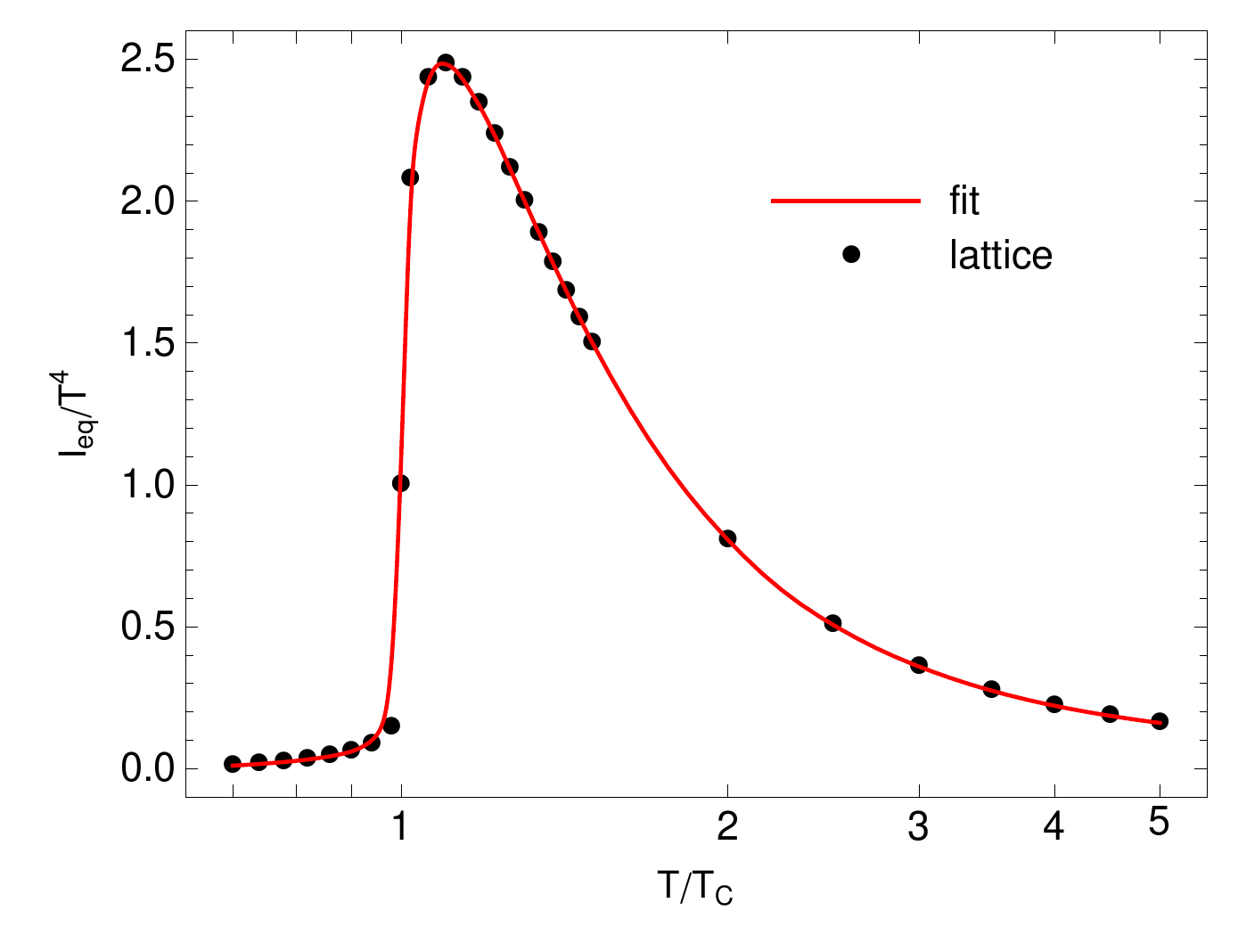}
\end{center}
\caption{(Color online)  Same as Fig.~\ref{fig:energy} but for the case of interaction measure. The lattice data are taken from~\cite{Borsanyi:2012ve}. }
\label{fig:trace}
\end{figure}
\begin{figure}[t]
\begin{center}
\includegraphics[angle=0,width=0.6\textwidth]{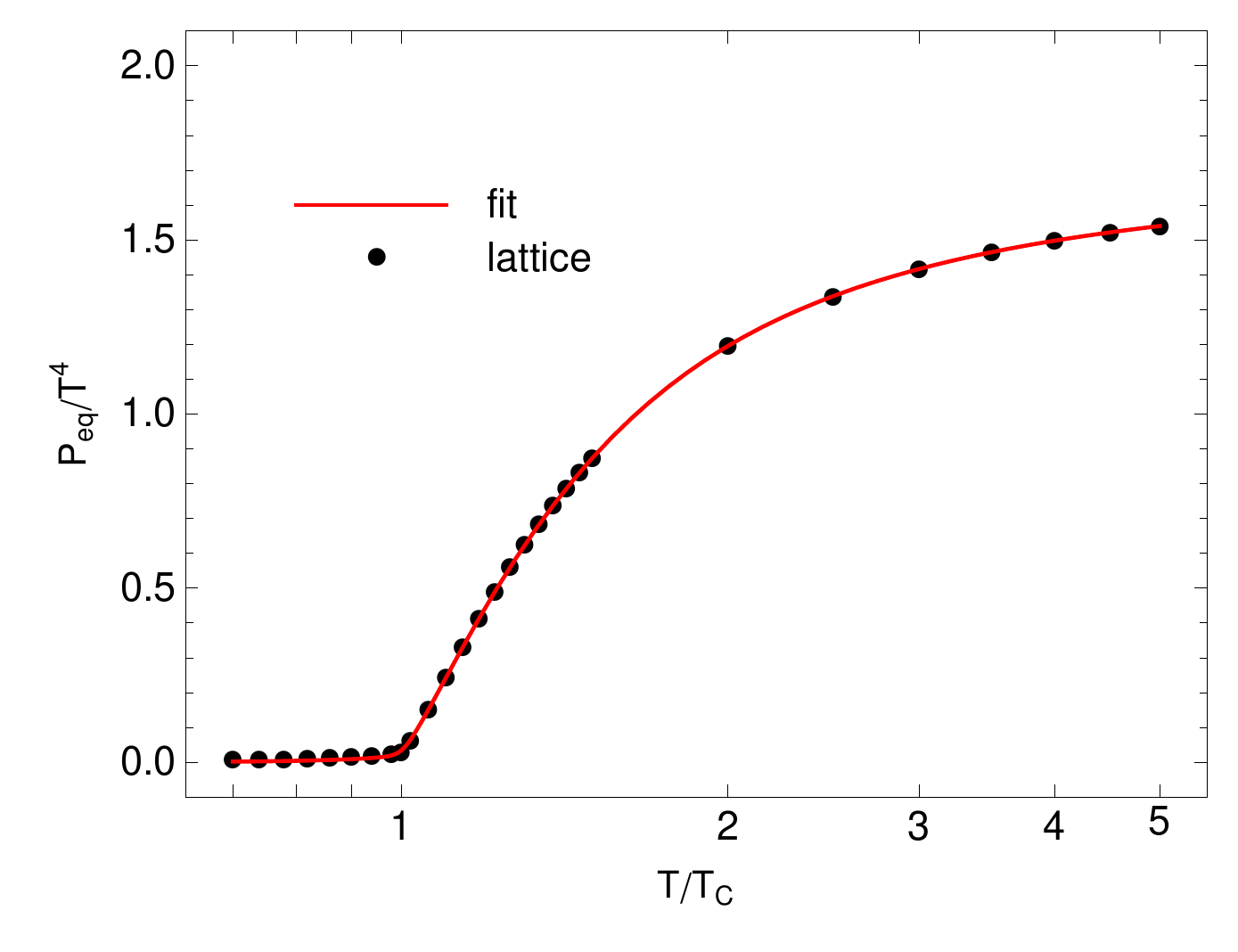}
\end{center}
\caption{(Color online) Same as Fig.~\ref{fig:energy} but for the case of pressure. The lattice data are taken from~\cite{Borsanyi:2012ve}. }
\label{fig:pressure}
\end{figure}

\subsection{Results for equilibrium} \label{sect:thermresults}

Our starting point is the analysis of the lattice results~\cite{Borsanyi:2012ve} for  the temperature dependence of the scaled trace anomaly $I_{\rm eq}(T)/T^4$. In order to guarantee that thermodynamic relations are satisfied exactly, we fit only $I_{\rm eq}(T)$. Subsequently,  using the resulting fit function $I^{\rm fit}_{\rm eq}(T)$,  the equilibrium pressure is calculated from Eq.~(2.3) given in Ref.~\cite{Borsanyi:2012ve}, namely
\begin{equation}
\frac{P^{\rm fit}_{\rm eq}(T)}{T^4}-\frac{P^{}_{\rm eq}(T_0)}{T_0^4}
=\int_{T_0}^T \frac{\dd T^\prime}{T^\prime}\frac{I^{\rm fit}_{\rm eq}(T^\prime)}{T^{\prime 4}} \, ,
\label{eq:P-func}
\end{equation}
where $P^{}_{\rm eq}(T_0=0.7 \,T_c)$ is read off again from Ref.~\cite{Borsanyi:2012ve}. We have checked that the temperature dependence of pressure resulting from (\ref{eq:P-func}) agrees with the lattice data at the level of 1--2\%.
Having $I^{\rm fit}_{\rm eq}(T)/T^4$ and $P^{\rm fit}_{\rm eq}(T)/T^4$, we calculate all remaining thermodynamic functions using the relations (\ref{eps1}) and (\ref{I}). 

Since $\seq$ is independent of the bag pressure, we use it to determine
$\gamma_\smallG(T)$. Our functional fit  to the lattice points describing entropy density vs. temperature is shown in Fig.\ref{fig:entropy} (solid red curve). The temperature dependence of the ratio $\seq/T^3$ shows a rapid
increase in the region of the phase transition, see Fig.~\ref{fig:entropy}. Such a behaviour
is usually attributed to the increase of the effective number of
degrees of freedom in the system.  Our model curve follows closely
this trend. We note that for the pure Yang-Mills theory studied
in~\cite{Borsanyi:2012ve} one expects a first order phase
transition, hence, the function $\seq/T^3$ should be discontinuous
at $T_c$. In our model calculation we rely only on  regular
approximations to the lattice data and, consequently, do not
reproduce exactly the character of the phase transition.

The temperature dependent Gribov parameter is shown in Fig.~\ref{fig:gamma}. The presence of the phase transition is reflected by a sudden decrease of $\gamma_\smallG$ in the region $T \approx T_c$. We note that for extremely high temperatures (not shown in the figure) the ratio $\gamma_\smallG/T$ becomes (to a good approximation) a constant. This means that the system may be treated approximately as conformal in this region. We shall come back to this point discussing the high temperature limit of the bulk viscosity in Sec.~\ref{sect:viscosities}.

Given the function $\gamma_\smallG(T)$ we find the temperature dependent bag pressure $B(T)$ from the identity
(\ref{Delta1}), which guarantees thermodynamic consistency\footnote{Knowing $\gamma_\smallG(T)$ one may equivalently use Eq.~(\ref{I}).}. The function $B(T)$ is shown in Fig.~\ref{fig:bag}. We have normalised $B(T)$ in such a way that it is zero at $T=0$. In the region of the phase transition $B(T)$ has a strong peak, while for large values of $T$ it becomes negative~\cite{Begun:2010md,Begun:2010eh}. We checked that at asymptotically large $T$ the bag pressure vanishes. In Figs.~\ref{fig:gamma} and
\ref{fig:bag} we show also the results obtained with the dispersion relation (\ref{EkM}).

Knowing the functions $\gamma_\smallG(T)$ and $B(T)$ we are in a position to calculate all other thermodynamic functions. In Figs.~\ref{fig:energy}, \ref{fig:trace} and \ref{fig:pressure} we show our results for $\epseq(T)/T^4$, $I_{\rm eq}(T)/T^4$ and $\Peq(T)/T^4$ (solid lines) compared to the lattice data~\cite{Borsanyi:2012ve} (dots). In all the cases we obtain a good agreement between our fits and the lattice data, also in the region of the phase transition.

\begin{figure}[t]
\begin{center}
\includegraphics[angle=0,width=0.6\textwidth]{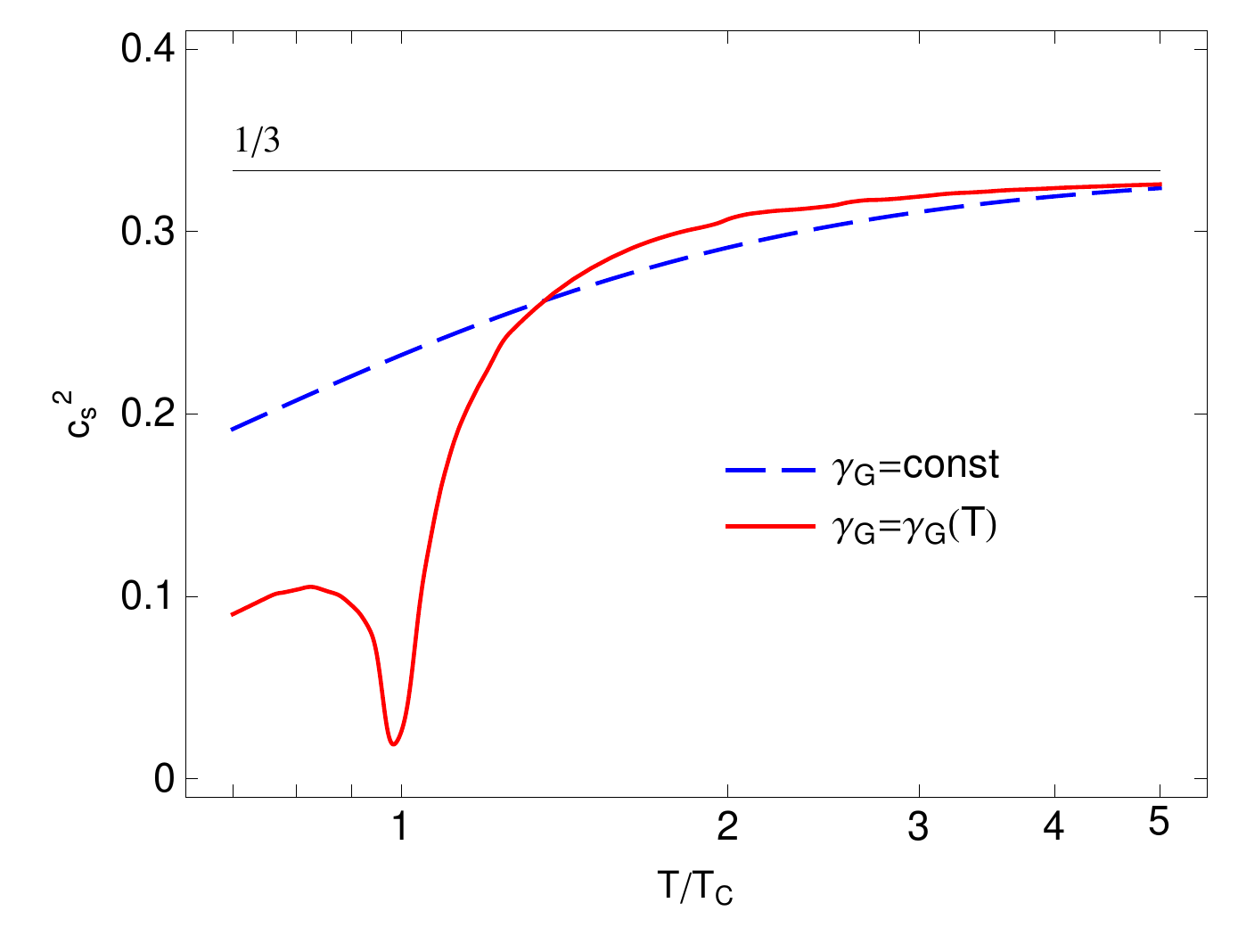}
\end{center}
\caption{(Color online)  Temperature dependence of the sound velocity squared for the temperature dependent (solid red line) and constant  (dashed blue line) Gribov parameter. }
\label{fig:cs2}
\end{figure}

Finally, in Fig.~\ref{fig:cs2} we show the sound velocity characterising the system (solid red line) that has been obtained from the formula
\begin{equation}
c_s^2 = \frac{\dd\Peq}{\dd\epseq} .
\label{cs2a}
\end{equation}
We note that the quantities $\Peq$ and $\epseq$ contain generally
contributions from the  bag pressure. They disappear if
$B~=$~const. The results for such a case,  with $\gamma_\smallG~=$~0.7~GeV,
are also shown in the figure (dashed blue line). We see that the
two results are consistent in the region above the phase
transition, where $\gamma_\smallG$ is approximately constant.

The temperature profile of the sound velocity has a dip typical for the phase transition (soft point). For the first order phase transition the sound velocity drops exactly to zero. Since we use only approximations to the lattice data, our result  for $c_s^2$ is finite at $T=T_c$ and reminds a rapid crossover transition~\cite{Chojnacki:2007jc}.

\section{Kinetics}\label{sect:kinetics}
\subsection{Lorentz covariance and boost invariance}
\label{sect:symetries}

Having discussed the equilibrium properties of our system we turn to the study of kinetic properties of the GZ plasma with the temperature dependent $\gamma_\smallG$ and $B$. To calculate the bulk and shear viscosities we consider a boost-invariant and transversally homogeneous system~\cite{Bjorken:1982qr}. These two properties facilitate our considerations --- they allow us to formulate a consistent (albeit simple) kinetic model within the relaxation time approximation used for the collision term. We note that although our treatment employs the  Bjorken scenario, the final results for the kinetic coefficients are independent of the expansion model, as they depend only on local thermodynamic parameters.

The boost invariance of our approach should be discussed in the context of the Lorentz invariance. In fact, the latter is explicitly broken by Eq.~(\ref{Ek}) that has been derived in the Coulomb gauge. A solution to this problem has been proposed in Refs.~\cite{Florkowski:2015rua,Florkowski:2015dmm}.  Since the Coulomb gauge as well as the value of the Gribov parameter are both determined by the procedures defined in the local rest frame of the fluid element, the energy of a gluon $E$ (denoting always the energy determined in LRF) should be treated as a scalar function of three-momentum. The details of this procedure are discussed in greater detail in Ref.~\cite{Florkowski:2015dmm}. Here we only explain our notation.

The longitudinal momentum of a particle is denoted by $k_\parallel$, the magnitude of the transverse momentum is $k_\perp \equiv \sqrt{k_x^2+k_y^2}$, and  $k^0$ is the magnitude of the three-vector ${\k}\equiv (k_x, k_y, k_\parallel)$, namely, $k^0 \equiv |{\k}|$. Thus, we use the four-vector $k^{\mu} = ( |{\k}|, k_x, k_y, k_\parallel)$ which has standard Lorentz transformation properties with $k \cdot k = k^2 = 0$. The function $E(\k, \gamma_\smallG)$ defines the energy of the gluon with three-momentum $\k$ in LRF. If the latter has the four-velocity $u$, the energy of a gluon can be written covariantly as
\begin{eqnarray}
E(k \cdot u,\gamma_\smallG) = \sqrt{ (k \cdot u)^2 + \frac{\gamma_\smallG^4}{(k \cdot u)^2} }.
\label{vareps}
\end{eqnarray}

For  one-dimensional, boost-invariant and transversally
homogeneous systems one may introduce the boost-invariant
variables: $\tau = \sqrt{t^2-z^2}$, $v =  k^0 t - k_\parallel z =
|{\k}| t - k_\parallel z$, $w = k_\parallel t -  k^0 z =
k_\parallel t -  |{\k}| z$~\cite{Bialas:1984wv,Bialas:1987en}. Hence, we find
\begin{eqnarray}
k \cdot u = \frac{v}{\tau} = \sqrt{\frac{w^2}{\tau^2} + k_\perp^2}
\end{eqnarray}
and
\begin{eqnarray}
E(\tau,w,k_\perp) = \sqrt{\frac{w^2}{\tau^2} + k_\perp^2  +
\frac{\gamma_\smallG^4}{\frac{w^2}{\tau^2} + k_\perp^2} } .
\label{Etau}
\end{eqnarray}
The phase space distribution function is a Lorentz scalar. This means that in our case the phase-space distribution function depends only on $\tau$, $w$, and $k_\perp$
\beq
f(x,p) = f(\tau,w,k_\perp). \label{f}
\eeq
Finally, we note that the Lorentz invariant measure in the momentum space expressed in the new variables is $\dd w\, \dd^2k_\perp/\tau$.

\subsection{Energy-momentum tensor and energy-momentum conservation
law} \label{sect:conserva}

As long as the system is in local equilibrium, its energy-momentum tensor has the perfect-fluid form
\begin{eqnarray}
T^{\mu\nu}_{\rm eq} = (\epseq + \Peq) u^\mu u^\nu - \Peq g^{\mu\nu},
\label{tmunu0}
\end{eqnarray}
where the flow vector  has the form $u^\mu = (t,0,0,z)/\tau$. For boost-invariant systems which are additionally homogeneous in the transverse plane, the general form of the energy-momentum tensor that holds also {\it out of equilibrium} is
\beq
T^{\mu\nu}\!=\!(\epstot+\PTtot) u^\mu u^\nu\!-\!\PTtot g^{\mu \nu} \!+\!(\PLtot - \PTtot) z^\mu z^\nu.
\label{Tmunu}
\eeq
Here $\epstot$ is the total energy density, $\PTtot$ is the total transverse pressure, $\PLtot$ is the total longitudinal pressure, and $z^\mu=(z,0,0,t)/\tau$. We note that the structure of  Eq.~(\ref{Tmunu}) follows directly from definitions of the components of $T^{\mu\nu}$. The total quantities include the particle and the field parts altogether, namely, we have
\beq
\epstot = \varepsilon +B, \quad \PLtot = P_\parallel -B, \quad \PTtot = P_\perp - B,
\label{tot}
\eeq
where~\cite{Florkowski:2013lya}
\begin{eqnarray}
\varepsilon &=& g\int \frac{\dd w\, \dd ^2 k_\perp}{ (2\pi)^3\, \tau} \, E(\tau,w,k_\perp) \, f, \label{epsandPs} \\
P_\parallel &=& g\int \frac{\dd w\, \dd ^2 k_\perp}{ (2\pi)^3\,  \tau} \,\frac{w^2 \left[1-\frac{\gamma_\smallG^4}{(w^2/\tau^2 + k_\perp^2)^2} \right]}{\tau^2 E(\tau,w,k_\perp)}  f, \label{ppar} \\
P_\perp &=& g\int \frac{\dd w\, \dd ^2 k_\perp}{ (2\pi)^3\, \tau} \,
\frac{k_\perp^2 \left[1-\frac{\gamma_\smallG^4}{(w^2/\tau^2 +
k_\perp^2)^2} \right]}{2\, E(\tau,w,k_\perp)}  f.
\end{eqnarray}
The standard pressure is then defined as $P = (2P_\perp+P_\parallel)/3$.

\medskip
In our case, the energy-momentum conservation law, $\partial_\mu T^{\mu\nu}=0$, simplifies to a~single equation
\begin{equation}
\frac{\dd \epstot}{\dd \tau} = - \frac{\epstot + \PLtot}{\tau}.
\label{encon1}
\end{equation}
Taking into account (\ref{tot}), this is equivalent to
\begin{equation}
\frac{\dd \varepsilon}{\dd \tau} + \frac{\dd B}{\dd \tau} = - \frac{\varepsilon + P_\parallel}{\tau}.
\label{encon2}
\end{equation}
Using (\ref{Etau}), (\ref{epsandPs}) and (\ref{ppar}) one
obtains from (\ref{encon2})
\begin{eqnarray}
 \frac{\dd B}{\dd \tau}
 &+ \,g \int \frac{\dd w \dd^2 k_\perp}{(2\pi)^3 \tau} \, \frac{2 \gamma_\smallG^3}{{(\frac{w^2}{\tau^2}+k_\perp^2)} E(\tau,w,k_\perp) }  \, \frac{\dd \gamma_\smallG}{\dd \tau} \, f
 \nonumber \\
 &+ \,g \int \frac{\dd w \dd ^2 k_\perp}{(2\pi)^3 \tau} \, E(\tau,w,k_\perp)\, \frac{\partial f}{\partial \tau}=0. \label{encon4}
\end{eqnarray}

\medskip
Before we analyse non-equilibrium features of Eq.~(\ref{encon4}), it is interesting to discuss the case of local equilibrium. In this case Eq.~(\ref{encon1}) can be written as
\begin{equation}
\frac{\dd \epseq}{\dd \tau} = - \frac{\epseq + \Peq}{\tau}.
\label{Bj1}
\end{equation}
Using thermodynamic identities $\dd \Peq = \seq \dd T$ and $\dd \epseq = T \dd \seq$ we obtain from (\ref{Bj1}) the Bjorken scaling solution for the entropy density~\cite{Bjorken:1982qr}
\begin{equation}
\seq(\tau)  = \frac{\tau_0\,\seq(\tau_0) }{\tau}.
\label{Bj2}
\end{equation}
This allows us to express the sound velocity of the system (\ref{cs2a}) by the formula
\begin{equation}
c_s^2 = \frac{\dd \Peq}{\dd \epseq} = \frac{\frac{\dd \Peq}{\dd T}}{\frac{\dd \epseq}{\dd T}}
= \frac{\seq}{T \frac{\dd \seq}{\dd T}} = - \frac{\dd  \ln T}{\dd  \ln \tau}.
\label{cs2b}
\end{equation}

\subsection{Mean field and kinetic equations} \label{sect:kineq}

If considered in equilibrium, the first two terms of
Eq.~(\ref{encon4}) agree with Eq.~(\ref{Delta1}). This suggests
that they may be treated as a non-equilibrium extension of
Eq.~(\ref{Delta1}). Therefore, as the first dynamic equation for a
non-equilibrium system we take
\begin{eqnarray}
g \int \frac{\dd w \dd ^2 k_\perp}{(2\pi)^3 \tau} \,
\frac{2 \gamma_\smallG^3}{{(\frac{w^2}{\tau^2}+k_\perp^2)} E(\tau,w,k_\perp) }  \, \frac{\dd \gamma_\smallG}{\dd \tau}
\, f +  \frac{\dd B}{\dd \tau}  = 0. \nonumber \\
\label{mfeq}
\end{eqnarray}
On the other hand, from the third term of
Eq.~(\ref{encon4}) we conclude that the distribution function may
satisfy the standard relaxation time approximation equation of the
form~\cite{Florkowski:2013lya,
Bhatnagar:1954zz,Baym:1984np,Baym:1985tna,Florkowski:2013lza}
\begin{eqnarray}
\frac{\partial f(\tau,w,k_\perp)}{\partial \tau}
= \frac{f_\GZ(\tau,w,k_\perp) - f(\tau,w,k_\perp)}{\tau_\GZ (\tau)},
\label{ke}
\end{eqnarray}
where
\begin{eqnarray}
\hspace{-1cm} \int \frac{\dd w \dd ^2 k_\perp}{\tau} \, E(\tau,w,k_\perp) \, f_\GZ(\tau,w,k_\perp) =
\int \frac{\dd w \dd ^2 k_\perp}{\tau} \, E(\tau,w,k_\perp) \, f(\tau,w,k_\perp).
\label{ec}
\end{eqnarray}
In Eq.~(\ref{ec}) we recognize the Landau matching condition for the energy. We note that Eq.~(\ref{ke}) has a formal solution \cite{Florkowski:2013lya}
\begin{eqnarray}
f(\tau,w,k_\perp) &=& f_0(w,k_\perp) D(\tau,\tau_0)  + \int_{\tau_0}^\tau \,
\frac{\dd \tau^\prime}{\tau_\GZ(\tau^\prime)} D(\tau,\tau^\prime)
f_\GZ(\tau^\prime,w,k_\perp),
\label{formsol}
\end{eqnarray}
with the damping function $D(\tau_2,\tau_1) $ defined as
\begin{eqnarray}
D(\tau_2,\tau_1)  = \exp\left[-\int_{\tau_1}^{\tau_2} \dd \tau^{\prime \prime} \tau^{-1}_\GZ(\tau^{\prime \prime}) \right].
\end{eqnarray}
The solution (\ref{formsol}) can be used to analyse the system's real off-equilibrium dynamics. However, in this work we concentrate on the kinetic coefficients solely, hence, it is sufficient to study Eq.~(\ref{ke}) in the linear response approximation. The latter applies for small deviations from local equilibrium.

\subsection{Bulk and shear viscosities} \label{sect:viscosities}

In order to determine the viscous corrections, we seek the solution of the kinetic equation (\ref{ke}) in the form
\begin{eqnarray}
f \approx f_\GZ + \tau_\GZ \, \delta f + \cdots \,.
\label{exp}
\end{eqnarray}
which gives
\beq
\hspace{-2.5cm} &&\delta f = - \frac{\dd \,f_\GZ}{\dd\tau}   \label{deltaf} \\
\hspace{-2.5cm} &&= -\frac{E}{T \tau}\left [ \frac{w^2}{E^2 \tau^2} \left(1\!-\!\frac{\gamma_\smallG^4}{(w^2/\tau^2 + k_\perp^2)^2} \right)
 \!+\!\frac{\tau \dd T}{T \dd\tau}
 \left(1\!-\!\frac{2 \gamma_\smallG^3 T}{\left(w^2/\tau^2+k_\perp^2 \right) E^2 } \frac{\dd\gamma_\smallG}{\dd T}  \right)
 \right]
f_\GZ \left(1+f_\GZ\right) . \nonumber
\eeq

From the definition of the bulk viscosity $\zeta = (P_\GZ-P)/\partial_\mu u^\mu$~\cite{Muronga:2003ta},  using Eq.~(\ref{deltaf}) and the Landau matching condition (\ref{ec}) in the linearised form,
\beq
\int \dd w \dd ^2k_\perp \,E\, \delta f = 0,
\eeq
we find
\beq
\hspace{-2.25cm}&& \zeta = -\frac{2 g \gamma_\smallG^4 \tau_\GZ}{3 T (2\pi)^3} \int \frac{\dd w \dd^2 k_\perp}{\tau (w^2/\tau^2 + k_\perp^2) }
 \label{zeta1} \\
\hspace{-2.25cm}&&
\times \left [ \frac{w^2}{E^2 \tau^2} \left(1 - \frac{\gamma_\smallG^4}{  (w^2/\tau^2 + k_\perp^2)^2}  \right) - \, c_s^2(T)
 \left(1 - \frac{2 \gamma_\smallG^3 T}{\left(w^2/\tau^2+k_\perp^2 \right) E^2 }  \frac{\dd \gamma_\smallG}{ \dd T}  \right)
\right]
f_\GZ \left(1+f_\GZ\right) . \nonumber
\eeq
Here we have used the Bjorken, leading-order solution and replaced the term $-\tau \dd T/(T \dd \tau)$ by the sound velocity squared, according to Eq.~(\ref{cs2b}). Changing integration variables: $w/\tau = k_\parallel$, $k_\parallel = k \cos\theta$,
$k_\perp = k \sin\theta$, and $y=k/\gamma_\smallG$, and integrating over $\theta$ from 0 to $\pi$,  Eq.~(\ref{zeta1}) can be cast into the form
\beq
\hspace{-2.25cm} \zeta = \frac{g \gamma_\smallG^5 \tau_\GZ}{3 \pi^2 T} \int_0^\infty \dd y\, \left[ c_s^2 \left( 1 - \frac{2}{y^4+1}
\frac{\dd\ln \gamma_\smallG}{\dd \ln T} \right) - \frac{1}{3} \frac{y^4-1}{y^4+1} \right]
f_\GZ \left(1+f_\GZ\right),
\label{zeta2}
\eeq
where the distribution function has now the form $f_\GZ = [\exp(\gamma_\smallG \sqrt{y^2+y^{-2}}/T)-1]^{-1}$. We note that for $\gamma_\smallG~=$~const., Eqs.~(\ref{zeta1}) and (\ref{zeta2}) agree with the formulas derived in Refs.~\cite{Florkowski:2015rua,Florkowski:2015dmm}.

It is interesting also to compare (\ref{zeta2}) with the expression for the bulk viscosity which is valid for the case where the dispersion relation (\ref{EkM}) is used. Following~\cite{Sasaki:2008fg} we find
\beq
\hspace{-2.25cm} \zeta_M = \frac{g M^5 \tau_{\rm rel}}{6 \pi^2 T} \int_0^\infty \dd y\, y^2\,
\left[ c_s^2 \left( 1 - \frac{1}{y^2+1}  \frac{\dd\ln M}{\dd\ln T} \right) - \frac{1}{3} \frac{y^2}{y^2+1} \right]
f  \left(1+f \right),
\label{zetaM}
\eeq
where $\tau_{\rm rel}$ is the appropriate relaxation time and $f = \left[\exp(M \sqrt{y^2+1}/T)-1\right]^{-1}$.

\begin{figure}[t]
\begin{center}
\includegraphics[angle=0,width=0.6\textwidth]{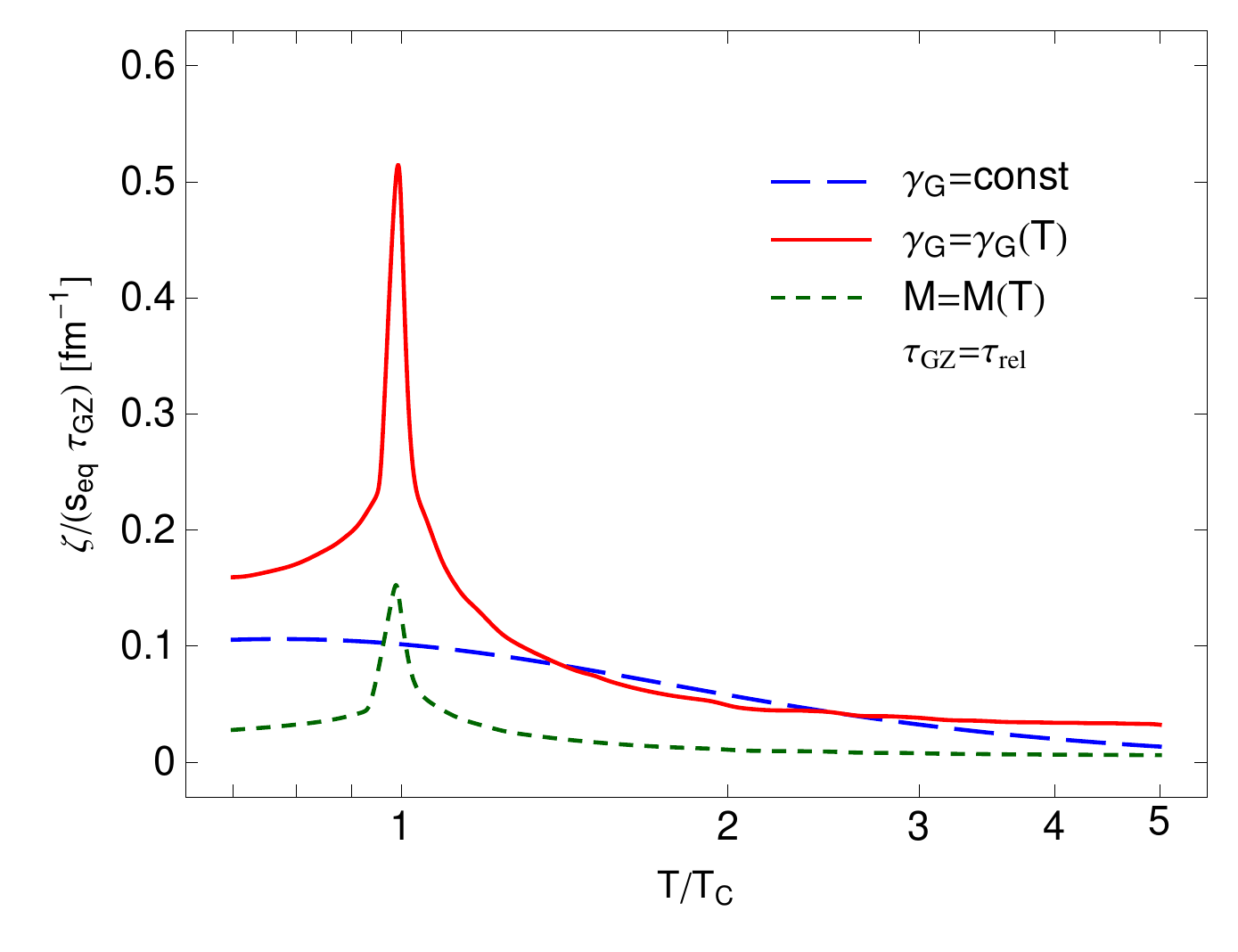}
\end{center}
\caption{(Color online)  Temperature dependence of the bulk
viscosity obtained from Eq.~(\ref{zeta2}) with the Gribov
parameter fitted to the lattice data (solid red line) and
with the constant value $\gamma_\smallG~=$~0.7~GeV (dashed blue
line). The short-dashed green line shows, for comparison, the bulk
viscosity obtained from Eq.~(\ref{zetaM}) with the effective mass
$M(T)$  fitted to the lattice data with $\tau_\GZ=\tau_{\rm rel}$.} \label{fig:bulk}
\end{figure}

Our results showing temperature dependence of the bulk viscosity (\ref{zeta2}) are presented in Fig.~\ref{fig:bulk} (solid red line). The most prominent feature of this behaviour is a sharp peak appearing in the region of the phase transition. The results obtained with the constant value  $\gamma_\smallG~=$~0.7~GeV~\cite{Florkowski:2015rua,Florkowski:2015dmm} (dashed blue line) agree with the present calculation in the region above the phase transition but they miss the peak structure. The latter appears in the calculation using the dispersion relation (\ref{EkM}) but its height and width are significantly smaller. 

It is interesting to observe that in the high temperature limit, where the Gribov parameter becomes proportional to temperature, $\gamma_\smallG = \alpha T$ with $\alpha = $~const., the bulk viscosity (\ref{zeta2}) vanishes.  Similar property has the bulk viscosity (\ref{zetaM}) if $M/T~=$~const.

\begin{figure}[t]
\begin{center}
\includegraphics[angle=0,width=0.6\textwidth]{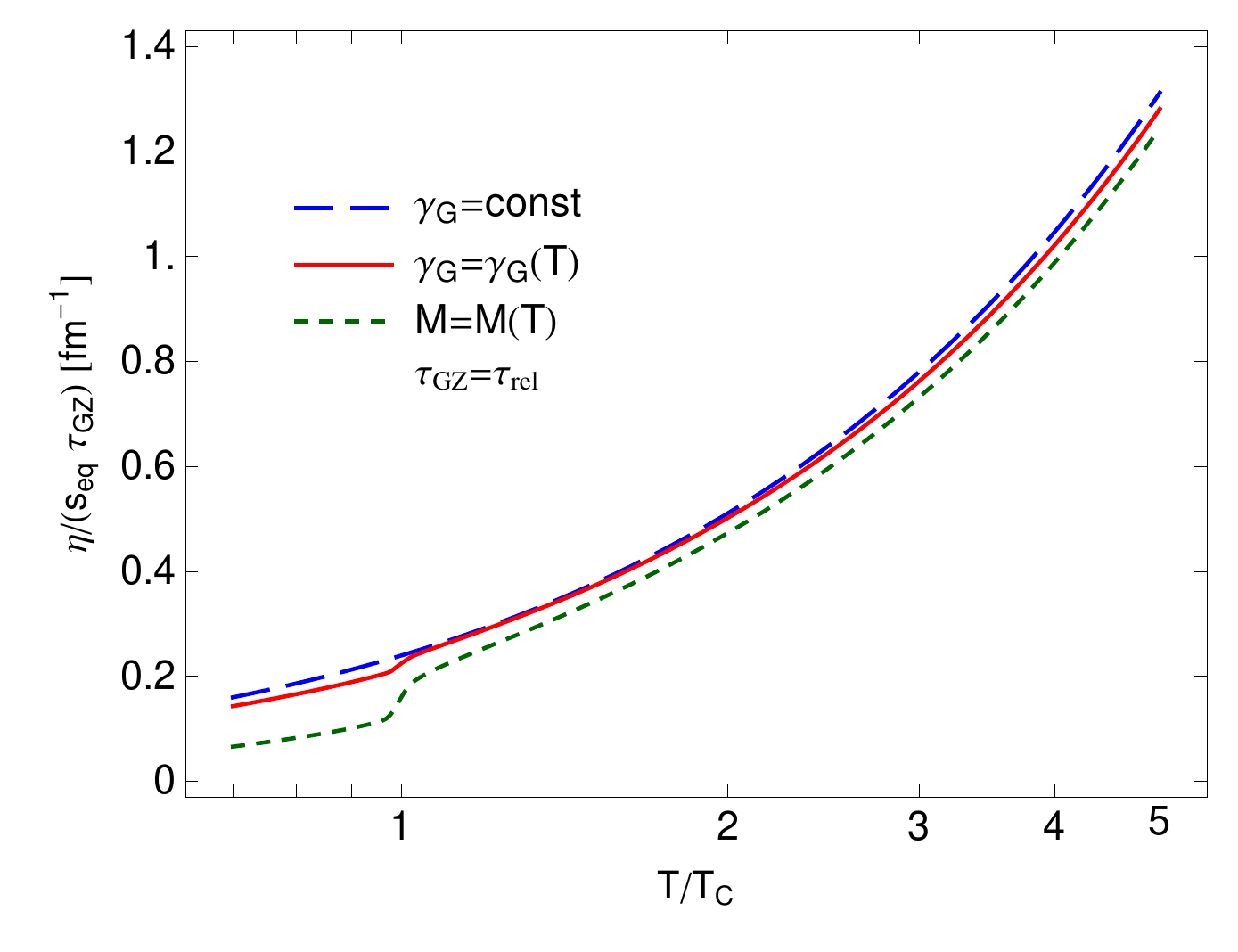}
\end{center}
\caption{(Color online) Temperature dependence of the shear viscosity obtained from Eq.~(\ref{eta2}) with the Gribov parameter fitted from the lattice data (solid red line) and with the constant value $\gamma_\smallG~=$~0.7~GeV (dashed blue line). The short-dashed green line shows for comparison the shear viscosity obtained from Eq.~(\ref{etaM}) with the effective mass $M(T)$ fitted from the lattice data and $\tau_\GZ=\tau_{\rm rel}$.}
\label{fig:shear}
\end{figure}

\medskip
The same method may be used to calculate the shear viscosity from
the relation $\eta = \onehalf (P_\perp
- P_\parallel)/\partial_\mu u^\mu$~\cite{Muronga:2003ta}. In this case we may keep only
anisotropic parts of $\delta f$, which leads to the expression
\beq
\hspace{-2.25cm} \eta =  \frac{g \tau_\GZ}{2 T} \int \frac{\dd w \dd ^2 k_\perp}{(2\pi)^3 \tau} \left(\frac{w^2}{\tau^2} - \frac{k_\perp^2}{2} \right)
 \left(1 - \frac{\gamma_\smallG^4}{  (w^2/\tau^2 + k_\perp^2)^2}  \right)^2 \frac{w^2}{E^2 \tau^2}
f_\GZ \left(1+f_\GZ\right).
\label{eta1}
\eeq
or, after changing the variables, to
\beq
\hspace{-2.25cm} \eta = \frac{g \gamma_\smallG^5 \tau_\GZ}{30 \pi^2 T} \int_0^\infty \dd  y\,  \frac{(y^4-1)^2}{y^4+1}
f_\GZ \left(1+f_\GZ\right).
\label{eta2}
\eeq
The corresponding formula for the dispersion relation (\ref{EkM}) is~\cite{Sasaki:2008fg}
\beq
\hspace{-2.25cm} \eta_M &= \frac{g_0 M^5 }{30\pi^2} \frac{\tau_{\rm rel}}{T} \int_0^\infty \dd y \, \frac{y^6}{y^2+1} f(1+f) \,,
\label{etaM}
\eeq
where  again $f = \left[\exp(M \sqrt{y^2+1}/T)-1\right]^{-1}$. The results showing the temperature dependence of the shear viscosity are shown in Fig.~\ref{fig:shear}. For three considered cases we find a similar monotonic behaviour. We note that we do not obtain the increase of the ratio $\eta/s$ for $T \to 0$. This feature is attributed to the chiral properties of pions at low temperatures~\cite{Csernai:2006zz}, which are missing in our approach.

\section{Conclusions}
\label{sect:conclusions}

In this work we have calculated the bulk and shear viscosity of
the Gribov-Zwanziger plasma in the case where the Gribov parameter
depends on temperature.  In this way we have generalised the
results obtained earlier in
Refs.~\cite{Florkowski:2015rua,Florkowski:2015dmm}. The
temperature dependence of the Gribov parameter has been determined
from the lattice data on the entropy density. The overall
thermodynamic consistency has been achieved by the use of
the temperature dependent bag pressure.

To calculate the kinetic coefficients we have introduced the
kinetic model based on the relaxation time approximation. We have
shown that this approach is consistent with the basic
energy-momentum conservation law. The final results for the
viscosities are proportional to the relaxation time, however,
otherwise, they depend only on the temperature of the system, in
particular,  via $\gamma_\smallG(T)$ and $B(T)$.  The evidence for 
a peaked structure of the bulk viscosity in the phase transition region
has been found. Compared to more standard calculations with a
temperature dependent mass, the use of the Gribov dispersion relation
yields a much broader enhancement.

\bigskip
{\bf Acknowledgments:} We thank Konrad Tywoniuk and Nan Su for
clarifying discussions concerning the Gribov approach to QCD. V.B.
and W.F. were supported by Polish National Science Center Grant
DEC-2012/06/A/ST2/00390. R.R. was supported by Polish National
Science Center Grant  No. DEC-2012/07/D/ST2/02125.

\end{document}